# Towards an Understanding of Why and How ICT Projects Are Initiated: Analysis via Repertory Grid

Htike Htike Wut Yi
*AUT University Private Bag 92006*
*Auckland 1142, New Zealand*
*htike.wutyi@gmail.com*

Stephen G. MacDonell[1]
*University of Otago*
*P.O. Box 56*
*Dunedin 9054, New Zealand*
*stephen.macdonell@otago.ac.nz*

## Abstract

*Contemporary business innovation relies increasingly on information and communications technology (ICT) solutions. As ICT initiatives are generally implemented via projects the management of ICT projects has come under increasing scrutiny. ICT projects continue to fail; as a result, while research in ICT project management has indeed increased, many challenges for research and practice remain.*

*Many studies have addressed the execution and management of ICT projects and the many factors that might relate to project outcomes. Very few, however, have considered ICT project initiation and the crucial decisions made at that very early, pre-life cycle stage. The primary intent of this research is therefore to investigate ICT projects with a particular focus on their initiation. In doing so we wished to understand why ICT projects are started, and how they are moved from idea or proposal to supported reality.*

*A combination of semi-structured interviews and the repertory grid data collection and analysis method was employed to investigate and validate the motivating factors that influence individual IT Managers' project initiation decisions and the methods they use to transition from idea to enacted project. Eighteen participants representing six medium-sized organizations were interviewed. A total of forty-nine recent ICT projects were identified and considered by these eighteen managers. A rich data set was collected and in-depth analysis was conducted. The results showed that there are indeed multiple underlying reasons for the decisions made at this early stage and that there are some especially common decision drivers. Some were expected, in the sense that they mapped to recommended best practice. For instance, most projects are motivated by a desire to achieve efficiencies or cost savings, and their potential tends to be assessed using cost benefit analysis. Other results were more surprising – competitor pressure was not a common driver for ICT project initiation in our analysis. Unsurprisingly, formal evaluation methods are more frequently used to assess project proposals when those projects are larger and higher profile.*

## 1. INTRODUCTION

The effective application of information and communication technologies (ICT) has become crucial to the operation of almost every organization, and the management of systems and technologies contributes, positively or negatively, to the very lifeblood of all businesses and other entities. Effective ICT management in both private and public sector organizations has become increasingly important due to highly competitive and time-constrained markets, the ongoing advancement of the underlying information and communication technologies, and larger scale changes such as the globalization of organizational activities.

In general, ICT development and adoption initiatives are implemented via projects (Cadle and Yeates, 2008); therefore, effective management of such projects plays a vital role in today's organizations. Not only are ICT projects often expensive, they also require substantial time and human resource commitment. The completion of these projects successfully, in a timely manner, within a specified budget, and meeting the users' requirements, is known to be problematic (Hillson, 2013). Beyond the technological challenges, also of influence are major contextual factors. The need for and impact of technology advancement and the adoption of emerging technologies and systems may vary in organizations with different cultural and political backgrounds. Therefore, ICT projects are collectively difficult to exemplify and theorize. Just as technologies evolve, the people who use ICT and systems directly or indirectly, and the organizational processes that constrain or are constrained by the development of these systems, must also be considered and managed. In undertaking ICT project management, then (referred to from this point as ITPM or IT project management), there are many visible and invisible factors to be taken into account, representing multiple scenarios of past, present and future use of technologies, systems and processes (Jani, 2008). In light of this, not surprisingly, ICT projects continue to fail globally at an alarming rate (Hillson, 2013). Initiatives that provide constructive insights into ITPM therefore have the potential to contribute significantly to organizational success. The research described in this paper is one such initiative. In particular, we investigate the very start of such projects, in an effort to understand why and how ICT projects are initiated.

[1] Corresponding author

## 2. ICT PROJECT STAGES

According to the Project Management Institute there are typically four main stages in projects, namely: initiation; planning and implementation; monitoring and controlling; and completion. The focus of the research reported here is on the first stage: initiation, wherein a project is suggested, assessed for viability and value, and progressed or discarded. Individual managers who initiate or propose a project are clearly influential in the decision-making process around that project. Generally, in medium- and larger-sized organizations, IT Managers (or those in similar roles, such as Chief Technology/Information Officers (CTOs or CIOs)) write ICT project proposals including requests for budget approval which necessarily highlight the need to implement such a project. A proposal normally includes a statement of the problem/opportunity, any recommended method(s) of addressing the problem/opportunity, and the anticipated benefits that would accrue from the expected outcomes of implementing the proposed project. Therefore, if the origin of a project is retraced, it should be possible to identify that it was initiated and implemented for certain reasons, with each 'go/no-go' decision based on potentially many factors.

If a project proposal is approved and signed off, the project is progressed towards implementation. Project team members are then expected to work to successfully deliver the project outcomes on time, on budget, meeting user requirements. Some projects are completed successfully while others face a complete or partial failure (Jani, 2008). In both successful and unsuccessful projects post-implementation reviews are generally conducted to assess whether, and to what extent, a project's objectives were met, to evaluate the effectiveness of the project approach, and to generate lessons learned. These reviews are often undertaken with an emphasis on project execution and outcomes (Cadle and Yeates, 2008), and so focus on benefits, effectiveness and efficiency, and constraints and issues encountered during the project.

However, looking more broadly, a failed project may be testimony to wrong or poor decisions being taken at the initiation stage. Thus, while aspects of a project's implementation will be important to that project's outcomes, the initiation stage and decisions made around it may be no less important. As stated above, every project is initiated for a reason, just as every decision made during the initiation stage is based on one or more motivating factors. If these early decisions are wrong or poor, subsequent project stages and outcomes may be at risk. Therefore, the underlying reasons that influence ITPM decisions during project initiation establish a foundation for that project. Most importantly, then, uncovering the 'why' factor at every decision point is needed for organizational learning. Some projects may be initiated based on decisions derived from self-belief and self-assessment of problems and opportunities, individual perceptions and experiences. In some instances, managers may make decisions subject to bias, consciously or unintentionally self-justifying the rationale for a project. In contrast, other projects may be initiated after the careful conduct of a range of formal processes and assessments. The intent here is to systematically investigate the rationale and methods that underpin decisions taken regarding ICT project initiatives. The aim is to explore how managers make decisions around project initiation and what factors drive or inform such decisions.

## 3. Related Work on Decision Making at ICT Project Initiation

ICT projects are often characterized as complex, requiring numerous decisions throughout the project life cycle (Sommerville, Cliff, Calinescu, Keen, Kelly, Kwiatkowska, McDermid, & Paige, 2012). To avoid later difficulties and to deliver a project successfully, the decisions made in the initiation stage are significant. However, researchers have seldom focused on the assessment of decisions around project initiation. Guah (2008) conducted a case study of IT project development in the UK National Health Service with a particular focus on how decisions are made in regard to IT projects. His findings noted that "...human decision-making is subjected to numerous biases, many of which operate at a subconscious level." (p.540). His findings further suggest that managers may engage in self-justification and may commit additional resources into projects even when the projects are actually poorly managed. In such instances, it is not uncommon that managers are unwilling to admit that their earlier decisions were wrong (Guah, 2008).

Shim, Chae and Lee (2009) note "...if a decision-maker's personal motivations are examined, a different explanation for risky decisions can be found." (p.1291). In some circumstances, decision-makers may be reluctant to explain or justify the rationale for the decisions made, or project managers may simply embrace projects without assessing the influencing factors at project initiation. That is, project managers are assigned to implement projects that have been approved and signed off. Therefore, the project manager's job is to execute the project and deliver the project outcomes successfully, not to assess the drivers of preliminary project decisions. According to Seiler, Lent, Pinkowska and Pinazza (2012, p.61), "motivation energizes and guides behavior toward reaching a particular goal and is intentional and directional." Thus, there are potentially numerous motivating factors and underlying reasons that influence project initiation decisions, a notion which has also been identified by Shim et al. (2009). They referred to the prior work of Keil, Wallace, Turk, Dixon-Randall and Nulden (2000), noting "...in cases of IT investment, decisions are more likely to be dependent on the decision-maker's intuition or personal motivation, because the formal decision-making process on IT investment is not well established in organizations." (p.1291). Thus, the decisions taken around project initiation may stem from a wide range of reasons and motivating factors, some of which may relate to the decision maker as much as they do to the project, the technologies and so on.

## 4. Research Objectives and Research Questions

Two particular aspects of ICT project initiation are probed here – why are projects undertaken, and how are project proposals 'moved forward' to become projects in reality. This study explores these issues in relation to a range of project types e.g., software selection; systems development, customization and implementation; outsourcing; technology selection; and the adoption of standards and frameworks. The work has the following primary objective: To explore the reasons underlying IT project initiation decisions and patterns of influencing factors and rationales.

The following research questions are posed in order to achieve the research objective:

- What factors drive ITPM initiation decisions and in which situations do IT managers initiate projects?
- Are there common patterns or significant differences of decision drivers across IT managers?

A second-level objective of the research is to contribute useful insights to the ICT practitioner community and to potentially enable better decisions to be made in the future. The collection of data from multiple organizations should support the discovery of patterns and exceptions regarding approaches used at the very beginning of projects.

## 5. Research Methods and Research Design

This research is conducted from an interpretivist foundation with the objective of the proposed work being pattern identification through interview analysis. It is an exploratory research endeavor, intended to reveal patterns in attitudes and opinions as well as commonly perceived problems and opportunities.

The repertory grid (RepGrid) technique was first developed by George Kelly in the 1950s in the context of psychological research and is an extension of Kelly's personal construct theory. Kelly originally introduced and applied the RepGrid technique in counselling his clients (Hunter & Beck, 2000). Kelly's psychology of personal constructs is conceptualized by Edwards, McDonald and Young (2009) as "...constructs are personal and ...may vary greatly among individuals. Fundamentally, a personal construct is an idea or concept that has been derived from specific experiences or instances of behaviour." (p.786). In Kelly's (1955) original work on personal construct theory, he reveals that "Man looks at his world through transparent patterns or templets [templates] which he creates and then attempts to fit over the realities of which the world is composed." [sic] (p.9). Kelly also describes that a person may predict an event in advance and then validate the construction as forecasted. Edwards, McDonald and Young (2009) summarized Kelly's personal construct psychology as follows: "Kelly believed that individuals act as scientists in order to understand their social surroundings: moreover, as people react with the world (and events occur) they continuously construct, amend and reform personal theories and assumptions. In other words, they build a model based upon experience that allows them to make predictions about future behaviour or interactions" (p.786).

The RepGrid method is an interview data collection and/or analysis technique that can enable a researcher to elicit personal constructs and to understand how individuals evaluate or construe the instance of a particular topic (Edwards, McDonald & Young, 2009). When used effectively the method can reduce the potential bias of the interviewer and affords flexibility to interviewees so that they are more able to describe their own interpretation of a specific topic (Hunter and Beck, 2000). The repertory grid method has been employed not only in its original psychological context but has also become popular in a number of study areas such as consumer research, marketing, nursing, clinical practice, management research and information systems.

In the last decade ICT researchers have increasingly utilized repertory grid methods in a variety of ways. Tan and Hunter (2002) in one of the earliest prominent works referred to the employment of repertory grid techniques in several previous, but more obscure, information systems research publications. Additionally, Tan and Hunter (2002) highlighted that understanding organizational cognition was becoming more important in IS research and they contended that the ignorance of IT professionals' cognition could impact on the outcomes of IS. They suggested the repertory grid method as highly recommended for the study of organizational and individual cognition in an IS context. Tan and Hunter (2002, p.40) noted: "This [repertory grid] technique offers the potential to significantly enhance our understanding of how users, managers, and IS professionals make sense of IT in their organisations."

The use of the repertory grid method in various ICT contexts can also be found in the following more recent studies. Rognerud and Hannay (2009) conducted research to identify the challenges in enterprise software integration in a major software development company, through the employment of repertory grids. Software practitioners' perceptions towards problem(s) encountered in this undertaking were elicited and analyzed. With regard to the integration project, the two alternatives were: either in-house software products will be integrated with each other, or third-party products will be integrated with the existing in-house products. Different perspectives and concerns of 'how' and 'what' to integrate had emerged in the company. In this study, important elements were elicited by asking participants about their most significant pain points and challenges, as well as their views on the pros and cons of software integration methods (known as constructs in RepGrid). After analyzing the grids and systematizing the different perspectives, Rognerud and Hannay were able to identify an optimal solution to the problem. The researchers were also able to present the results in a company seminar specifying courses of action for the current, on-going and future integration projects. Thus, as can be concluded from Rognerud and Hannay's application of repertory grid in this particular study, the RepGrid method not only assists with capturing different perceptions and opinions but it can also aid in potentially disparate or divided groups arriving at an acceptable solution.

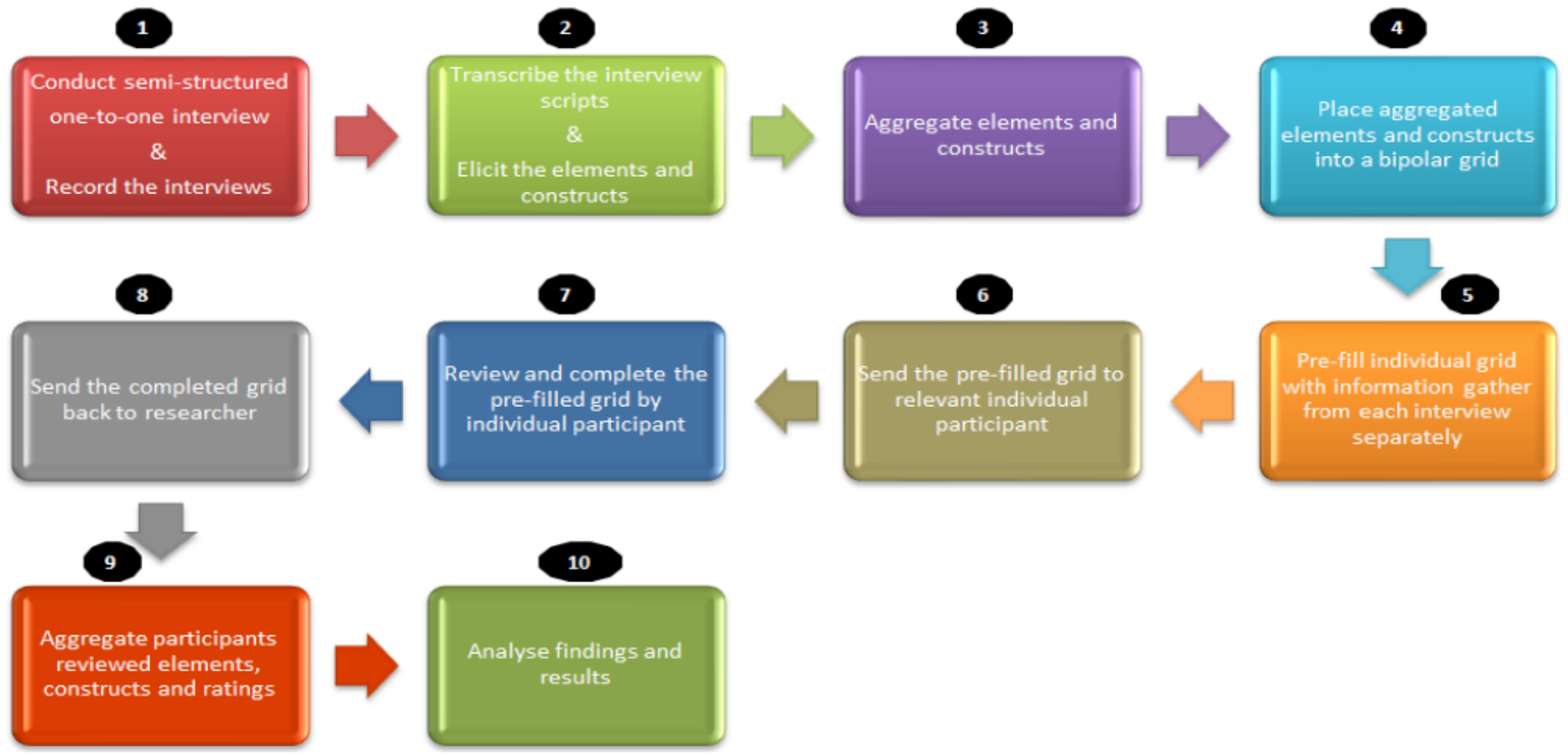

Figure 1. The process flow of the customized Repertory Grid Method use in this study

Employment of the repertory grid method can also be seen in Siau, Tan and Sheng's (2010) empirical study that had the objective of identifying the important characteristics of software development team members. With the assistance of the RepGrid method, the outcomes of their study not only achieved their research objective but practical guidelines for human resource allocation and development training requirements for IT practitioners (particularly in development teams) could also be generated. Siau et al. (2010) encouraged the wider use of the repertory grid method in other information systems research. The method was also adopted by Napier, Keil and Tan (2009) in their study of IT project managers' construction of successful project management practice, as well as Hunter and Beck's (2000) research in cross-cultural information systems.

The repertory grid method is adopted in this research project because it focuses on people, their understandings and how they construct their view of the world. Our main research interest lies in the drivers of IT managers' decisions at the project initiation stage with the primary objective 'To explore the reasons underlying IT project initiation decisions and patterns of influencing factors and rationales'. Individuals use cognition capabilities and personal constructs (Kelly, 1955) when making decisions. Therefore, given an emphasis on IT managers' decisions based on their interpretations, assumptions and experiences, the repertory grid method was seen as being particularly suitable for use in this study.

A tailor-made repertory grid method was employed in order to balance the analytic strengths of the RepGrid method with the time required in data collection. In this study, an alternative means of collecting the data one-on-one from multiple managers in a suitable timeframe was designed and implemented. This was done in consideration of participants' availability and to minimize the potentially lengthy data collection process while maintaining the intention of obtaining a rich and in-depth data set that genuinely reflected participants' project initiation experiences and perceptions. The characteristics of the customized repertory grid method used in this research can be seen in the ten-step process flow diagram shown in Figure 1.

Forty-five invitations were sent to IT professionals from ten medium- and large-sized organizations. These IT professionals deliberately included project sponsors, project managers, enterprise architects, solution architects, business systems managers and CIO/ICT directors. It was expected that approximately 30% of the invitees would accept the request to take part, leading to a likely sample size of 15 participants. According to Tan and Hunter (2002), "A sample size of 15 to 25 within a population will frequently generate sufficient constructs to approximate the universe of meaning regarding a given domain of discourse." (p.50).

A total of 21 participants from six medium and large organizations accepted the invitations. The organizations were two commercial banks, two not-for-profit organizations and two transportation service companies. Out of the 21 participants, three were unable to identify relevant projects due to their short-term tenure at their current company at that time. Therefore, a total of 18 respondents participated in the research project. The 18 participants, representing six medium and large organizations, consisted of one project sponsor, three ICT directors (CIOs), two solution and enterprise architects, one analyst, and eleven IT managers including business systems managers and project managers. Therefore, the preliminary assumption of an interview acceptance rate of around 30% proved to be approximately correct. All participants were interviewed individually at the participant's choice of place, date and time. The 18 interviewees identified a total of 49 IT/IS projects in which they either held leadership roles or were party to the decision-making of the project teams.

First, participants were interviewed individually (by the first author) using a semi-structured interview technique.

Table 1. Summary of results presentation structure

| Section | Analysis | Description |
|---|---|---|
| **6.1** | **Why** | Reasons, problems or opportunities that encouraged or led to the project being undertaken |
| **6.2** | **How** | The approaches used in going from the idea, problem or opportunity to the solution |
| **6.3** | | Reasons for using a specific approach |

The interviewer did not ask any leading questions; however, general prompts were used to ensure that the experiences and actions the interviewees reported were related to the topic and were within the subject matter of the research (Rognerud & Hannay, 2009). During the interview, the interviewer acted as a facilitator and participants (the interviewees) defined their own elements and constructs (Alexander, Loggerenberg, Lotriet & Phahlamohlaka, 2010). They first listed the topics (i.e., one to five projects) that they had been involved in and wanted to describe for the interview. Then, the content covered within the topic was also chosen by the interviewees. They were not interrupted unless the conversation moved in a direction beyond the research focus on IT project management decisions during project initiation. The entire conversation was audio recorded with the participants' consent. The conversations were then transcribed and elements and constructs from each interview were elicited by the first author. All elements and constructs were aggregated and constructs were placed into a bipolar grid. One side of the poles of a construct represents elicited constructs while the other side of the poles reflects contrast opposites. The grid containing aggregated elements and constructs was then sent to each participant. In the aggregated grid, the intermediate outputs gathered from each interview (e.g., participant information, project names, factors considered) were pre-filled and sent to the relevant participant in order to obtain confirmation and reassurance of the information that the participants provided at interview. Furthermore, participants were also able to rate the constructs they considered relevant and applicable to the projects that they had identified.

## 6. Results and Discussion

This section represents the core of the contribution emanating from the research, which probed two specific aspects of project decision-making at initiation – specifically, why projects are undertaken, and how projects are moved from proposal to reality. To present multiple perspectives on such factors, the interview results from IT managers, project managers and sponsors from the six organizations are categorized, summarized and analyzed. As described above, different types of respondents actively participated in the interview process. In order to facilitate concise descriptions, the generic term 'IT Manager' is used to represent the entire population of participants.

It should also be noted that the participants' responses are regarded as being reflective of their views at the time project decisions were being made. An assumption is also made that IT Managers' descriptions of the underlying reasons and influencing factors around project initiation convey their bona fide interpretation of events. It does not necessarily mean that respondents' opinions at the time of project initiation reflect their current perspectives or mental models. Therefore, an unintentional commentary might emerge in the reminiscences of a past decision-making process, based on participants' experiences in the interim. However, in order to capture the most accurate picture possible, participants were asked to consider their most recent projects and were reminded to try to provide a 'snapshot' of reasons and influencing factors that were relevant at the time of project initiation. In addition, and as described above, a two-step process was used in data collection. After the interview process the participants were supplied with the aggregated grid, so respondents were given another opportunity to recollect (and potentially revise) their narration. Therefore, it is believed that the collected data reflect the participants' actual intended responses regarding their identified projects.

Results are presented with an emphasis on the decision-making aspects; specific approaches, methodologies and frameworks are not explained in detail, based on an assumption that readers are familiar with the terminologies and processes commonly referred to in ICT projects (e.g., use cases, prototypes, SaaS, bespoke development, requests for proposal). Table 1 summarizes each section and sets out how the analyses of results are presented.

Each section is subdivided into two sub-sections, namely, Findings and Discussion. In the Findings sub-section the results are reported. Specific findings are then elaborated with narration in the Discussions sub-section. In sections 6.1 and 6.2 simple summary percentages are presented. For instance, 45% of projects were initiated with one of the primary reasons being to obtain cost savings. In section 6.3, a summarized grid is presented for each element. In other words, the selected approach (element) is illustrated with its summarized rating scale. Participants were asked to rate the reasons (constructs) on a scale of 1 to 5 in relation to each selected element. If the construct at the left-hand side of the grid most accurately represented their reason for selecting an approach or method, a rating of 1 would be given. In a similar way, a rating of 5 would be chosen if the right hand grid construct most defined their rationale. A rating scale value of 3 would indicate that both rationales were equally applicable.

For instance, an IT Manager might select a packaged solution (element) for a particular project due to their perception of both faster development and long term benefits, which are on the left and right poles of the grid, respectively; a rating of 3 would be appropriate. When the grids are aggregated, if there are 5 projects with the same rating scale value of 3, the total count of 5 will be displayed in the aggregated grid at the relevant cell. As illustrated in Figure 2, each approach (element) is considered in a separate aggregated repertory grid with its associated series of rationales (constructs) listed in left hand and right hand poles. The number of times participants rated each scale value against an individual construct are summed and placed in the corresponding cell. (Further elaboration of the content of the grids and how the individual repertory grids are consolidated is explained in detail using actual ratings in sub-section 6.3.)

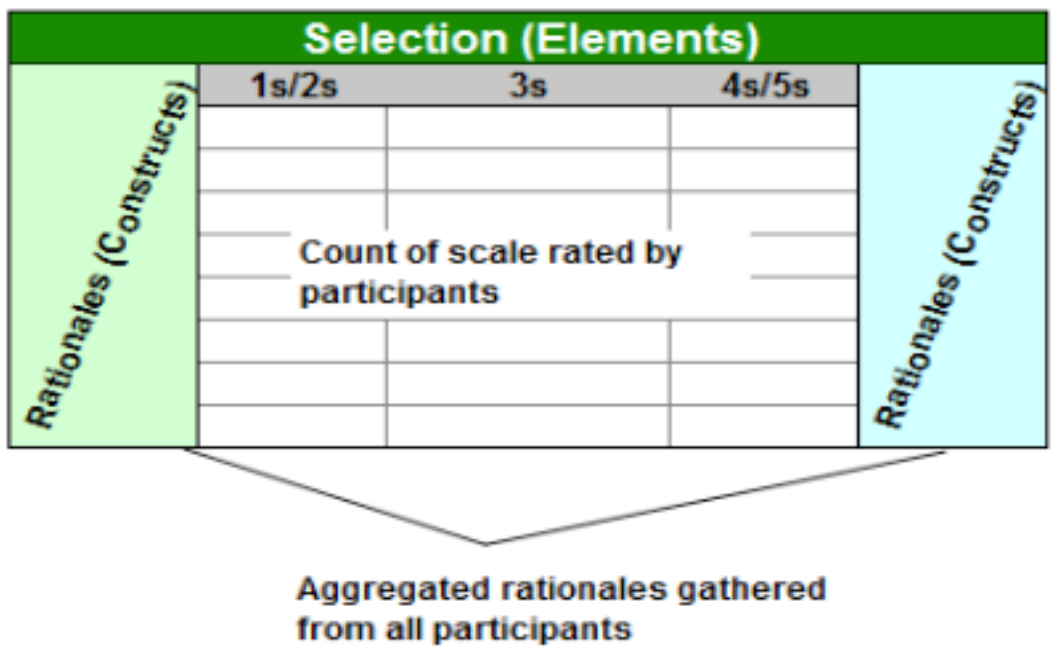

| Rating Scale | Description |
|---|---|
| 1s/2s | The rationale from ***left*** hand column match with the rationale for a particular project |
| 4s/5s | The rationale from ***right*** hand column match with the rationale for a particular project |
| 3s | The rationale from ***both left and right*** hand column match with the rationale for a particular project |

Figure 2. Description of summarized repertory grid results

### 6.1 Why Are ICT Projects Undertaken?

The 'Why' question was asked first to determine managers' preliminary rationales in the early stages of project initiation. In many respects, these 'Why' factors relate to the pre-initiation stage of a project, and indicate the main motivating factors and justification for projects being initiated.

***Findings***

A total of 36 underlying reasons were identified by the 18 participants during the interview process. When the aggregated grid was distributed to the respondents, the rationales were consolidated to 34 as shown in Table 2.

Among the 34 rationales, 'Inefficient system/process', 'Cost savings' and 'Process improvement/change' are the top three motivators for project initiation. In total, 47% of projects were initiated due to systems or processes perceived as being inefficient. The second top trigger for projects being initiated was 'Cost savings', at 45%. The third most common motivating factor for project establishment was 'For process improvement or change'. Of note in the other responses is the finding that only 14% of projects were being undertaken in response to a business case.

During the interview process, the primary drivers of two projects were identified as being due to political reasons. However, when the participants were provided with the aggregated repertory grids, the 'Political reasons' rationale was not selected. Similarly, one of the interviewees initially identified that a project had been initiated to convince top management of the potential of a particular solution; however, this reason was not included when the repertory grid was returned by that interviewee. In this respect there are discrepancies between the interview conversation outcomes and the repertory grid returns. It may be that participants did not see these reasons as primary drivers of project initiation when they contemplated the past events (although participants were encouraged to select all drivers/factors that applied). On the other hand, it may be that participants expressed unhesitatingly or felt more comfortable during the initial face-to- face conversation rather than when asked to more formally record their views by way of completing the grid. Note that efforts were made to avoid such limitations – the participants were advised that the grids were only being utilized to ensure all the analyzed data were genuine reflections of participants' views rather than the researchers' (incorrect) interpretation of past events and situations. Apart from a small number of such discrepancies, the majority of the interview results and repertory grid returns matched. What is more, a richer data set was obtained as participants had an opportunity to reflect on their project reasons/rationales while also considering others' responses.

***Discussion***

The rationales for projects being undertaken are many and varied. Some may be based on a business case constructed according to an organization's predefined process, and executives may approve it as long as the business case shows 'fitness-for-purpose'. However, the more in-depth analysis of potential consequences, latent contingencies and residual risks of projects being implemented may be deficient in many organizations. Of course, it is difficult to anticipate future problems especially in the context of rapid technology innovation. However, at the very least, thorough consideration of pre-project assessments should be made in terms of how a new project would be integrated with or fit into existing systems, structures, processes and other initiatives.

In most cases, project managers (PM) are appointed only after the initiative has been approved (Sauer, Gemino and Reich, 2007). Due to the nature of project assignments to project managers, the PM's role is to successfully deliver the project as per the business case, regardless of the validity of the underlying reasons. What is more, PMs may be asked or required to continue with failing projects (Jani, 2008). Some projects may not be suitable in a particular organization environment/culture at a specific point in time. For example, due to a consequent significant culture shift, extensive change management procedures may need to be implemented as a pre- project assignment before the actual IT project is initiated. Some may argue that one of the critical factors for IT project failure is user resistance. However, if business cases were to include a distinction between business risks, technology risks and project risks with in-depth analysis, the PMs may be better able to plan with more accurate project estimation and execute projects with better change management control. This does not necessarily mean that an in-depth analysis and justification at the initiation stage will lead to project success; to a certain extent, however, it would support the delivery of a successful project. Business cases can enable informed decisions to be made on proposed resource consumption in terms of effort and budget along with risk assessment, cost benefit analysis and alternative solution analysis.

Table 2. Summary of 'Why' results

| What were the problems or opportunities that encouraged or led to the project being undertaken? | Grand Total | Percentage for a reason |
|---|---|---|
| Inefficient system/process | 23 | 47% |
| Cost savings | 22 | 45% |
| Process improvement/change | 21 | 43% |
| Management/company expectations | 19 | 39% |
| Unsupported legacy system | 17 | 35% |
| New business/Techlogy opportunity | 17 | 35% |
| Reduce manual process | 16 | 33% |
| BAU improvement | 14 | 29% |
| Fragmented systems | 13 | 27% |
| Cope with growth | 12 | 24% |
| Benefit realisation | 12 | 24% |
| Integrated processing/reporting challenges | 11 | 22% |
| Customer focus | 11 | 22% |
| Other organisations are applying the particular technology successfully | 10 | 20% |
| Better management control | 10 | 20% |
| Lack of standard process | 9 | 18% |
| Business case | 7 | 14% |
| IT Infrastructure changes | 7 | 14% |
| Data integrity issues | 6 | 12% |
| All digital | 6 | 12% |
| Existing techlogy issues | 6 | 12% |
| To increase market share | 6 | 12% |
| Reduce support needs | 5 | 10% |
| Compliance to internal/regulatory requirements | 4 | 8% |
| Competitor pressure | 4 | 8% |
| Concurrent upgrade/customisation | 4 | 8% |
| Company's strategic plan changes | 4 | 8% |
| Staff reduction | 3 | 6% |
| Develop in-house capability | 3 | 6% |
| Customer complaint | 3 | 6% |
| Public relations opportunity | 3 | 6% |
| Departmental or team invation | 2 | 4% |
| Executive management's expectations | 2 | 4% |
| Organisational change | 2 | 4% |
| Political reasons | 0 | 0% |
| To convince top management | 0 | 0% |
| Other(s) (please sepcify in the cell alongside) | 0 | 0% |

As is implied in the above 'Findings' sub-section, participants typically identified at least two influencing factors per project. Among those, only 7 projects (14% of the total projects) were initiated based on a business case. This seems to be in conflict with perceived best practice. As Cadle and Yeates (2008) argued, "No project should be undertaken without first establishing a business case for it – without, in other words, showing that it is justified. The business case defines what is to be done, why, and what are the timescales and costs involved" (p.31). Also, during the interview process, one of the participants argued strongly that any IT project must be initiated with a business case and a new technology/system should not be introduced without a complete and comprehensive business case.

### 6.2 How Did You Move To A Solution?

After the participants provided the preliminary reasons for initiating their nominated projects, the 'How' question of 'What approaches did you use in going from the idea, problem or opportunity to the solution? How did you move towards a solution?' was put to them. The intent was to investigate the methods or approaches participants elected to use to implement their ideas; in other words, their chosen approach(es) for moving their project ideas to a supported reality.

#### *Findings*

A total of 19 approaches were identified by the 18 participants during the interview process. When the summarized grid was distributed and the participants returned the reviewed grids, the approaches or methods were selected as shown in Table 3. As the table illustrates, IT Managers have primarily undertaken cost benefit analysis and requirements gathering/specification and analysis to move their projects forward. In contrast, site visits appear to be used relatively rarely as does the RFI process.

#### *Discussion*

When organizations intend to undertake IT projects there is generally a sequence of processes employing a number of approaches/methods that will be carried out. However, depending on the culture of those organizations and business units, and the experiences and expertise of individual IT Managers, the types of approaches/methods used may vary. As can be seen from the above results, it appears that a requirements gathering, specification and analysis process was conducted in relation to around half of the projects considered. This was a lower than expected result given evidence in the research literature. Without systematically identifying, gathering, specifying and analyzing the requirements (Aurum and Wohlin, 2005), an understanding of stakeholder needs and customer/user expectations may not be able to be established. A lack of requirements understanding and incomplete or changing requirements are frequently noted as among the critical factors in project failure (Hansen, Berente and Lyytinen, 2009). One of the participants noted, "without capturing the requirements first, our requirements could end up what software providers can offer and what they demonstrate. So, we decided not to go out to the market and not to submit RFP until we've got a complete set of requirements."

Table 3. Summary of 'How' results

| What approaches did you use in going from the idea, problem or opportunity to the solution? (How did you move towards a solution?) | Grand Total | Percentage for a reason |
|---|---|---|
| Cost benefit analysis | 28 | 57% |
| Requirements gathering/specification and analysis | 26 | 53% |
| Project Management methodology | 21 | 43% |
| Internal organisations discussions | 21 | 43% |
| Vendor's demos | 19 | 39% |
| Market research | 16 | 33% |
| Request for Proposal (RFP) process | 14 | 29% |
| Company's predefined process | 7 | 14% |
| Evaluation framework | 6 | 12% |
| Prototype | 6 | 12% |
| Models (e.g. use cases) | 6 | 12% |
| Narrative specification | 5 | 10% |
| Request for Information (RFI) process | 4 | 8% |
| Site visits | 3 | 6% |
| No Project Management methodology applied | 3 | 6% |
| Other(s) | 3 | 6% |
| No specific approach applied | 1 | 2% |
| Request for Tender (RFT) process | 0 | 0% |

Another result of note is the participants' limited use of 'Market research', employed in only a third of the projects. On referring back to the individual grids, it appears that IT Managers either researched the market or conducted 'Request for Information/Proposal' (RFI & RFP) processes, while a few did not conduct any of these activities. During the interview process one IT Manager expressed the advantages they gained during the decision making process on the basis of site visits. However, as noted briefly above, the 'Site visits' activity does not seem to be a particularly common practice.

It should be noted here that there were a few participants who were from the same organization and they unanimously indicated that their project processes were driven by their company's pre- defined standard approaches. For this reason, their approaches and methods were derived purely from their organization, and were not based on an individual IT Manager's project management style or preferences.

### 6.3 Why Did You Use Certain Methods To Move Towards A Solution?

In this section, the reasons for selecting the particular approaches, processes and methods used are addressed. During the interview, participants were requested to provide the rationales behind their chosen approach(es). Each individual participant's answers were placed in a bipolar grid and the pre-filled grid was sent back to the relevant participants through email communication. The participants then reviewed, edited and rated the pre-filled grid and sent the finalised grid back to the researchers. All the participants' returned grids are consolidated into the single grids that are considered in the following sub-sections.

#### *Findings*

In each consolidated grid, each element (approach, process, method) that participants identified is placed at the top and the contrast constructs are placed at the left and right hand side of the grid. The ratings are counted and the total frequency count of ratings is placed inside the grid. For instance, the following explanation conveys the elaboration of Table 4. (Specific issues of note or those selected for discussion are indicated on each grid using circles or rectangles.)

- **Element** = 'Vendor's Demos' (i.e., the 'How' factor; the chosen approach);
- **Construct** = 'Faster development/implementation process' & 'Quality focus' (i.e., the 'How-Why' factor; the reasons for selecting the 'Vendor's Demos' approach);
- **Rating** = 1s/2s, 3s, 5s/4s; rating of 1s and 2s = the participant's reason for selecting the particular approach matches the left hand pole construct; rating of 5s and 4s = the participant's reason for selecting the particular approach is defined by the right hand pole construct; rating of 3s = both left hand and right hand constructs are equally applicable. For example, if a participant believed that a reason for using 'Vendor's Demos' was that it resulted in faster development or implementation, they will select a rating of 1 or 2 depending on the intensity. However, if a reason for their selecting the 'Vendor's Demos' approach is due to a quality focus, a rating of 4 or 5 will be given depending on the strength of significance. If selecting such an approach was based on a mix of both expected faster development and a quality focus, a rating of 3 will be selected. In the figure, the total count of 2 under 1s/2s indicates that two IT Managers used the 'Vendor's Demos' approach due to a perception that it would lead to a ‘Faster development/implementation process’. However, three IT Managers employed the ‘Vendor’s Demos’ approach due to its ‘Quality focus’ (under 5s/4s). The count of 1 for rating 3s indicates that one IT Manager believed ‘Vendor’s Demos’ can result in a faster development process but is equally quality focused.

that IT Managers also believed that the requirements gathering approach is not easy to apply or use and is reliant on having the right people available.

As can be seen in Table 6, the non-use of a project management methodology is not a common occurrence. Not surprisingly, the results suggest that this approach

Table 4. Analysis of ‘How-Why’ results; ‘Vendor’s Demos’ element

| Vendor's Demos | | | | |
|---|---|---|---|---|
| | 1s/2s | 3s | 5s/4s | |
| Faster development/implementation process | 2 | 1 | 3 | Quality focus |
| Clear responsibilities for all parties | 0 | 1 | 7 | Easy to use |
| Uncertainty of the solution/method | 4 | 3 | 1 | Availability of an instant solution |
| Time and/or budget constraints | 2 | 1 | 1 | Specified in process |
| Problem focused | 1 | 0 | 6 | Solution focused |
| Minimal risks | 1 | 3 | 0 | Governance procedures |
| Useful for communication | 2 | 4 | 2 | Useful for implementation |
| Small/low-impact project | 0 | 1 | 5 | Large/ high profile project |
| Easy to apply | 0 | 3 | 2 | Can be verified |
| Company's standard approach | 3 | 2 | 1 | Personal or team preference |
| The most common/acceptable practice in IT industry | 4 | 2 | 2 | A tailor-made process based on the project |
| Human resources constraints | 0 | 3 | 2 | Availability of internal staff/external consultant |
| Regulatory compliance | 0 | 1 | 5 | Freedom to choose |

The above findings suggest that some IT Managers regarded the vendor’s demonstration as supporting a faster development/implementation process while others considered it to support a quality focus. Notably, all IT Managers who used this approach supported it as being easy to use. It appears that some IT Managers utilized a vendor’s demonstration when they were uncertain of the solution or method whereas others might have reviewed and appraised vendor’s demos for reassurance purposes. It was considered as being a useful method for both communication and implementation. As can also be seen, the total count of 5 under the 5s/4s column for ‘Large/high profile project’ suggests that vendor’s demos were used principally when the projects were considered as large and/or high profile. Additionally, IT Managers not only regarded the approach as the most common and acceptable practice in the IT industry but they also believed that it enabled them to select from many alternatives.

should only be used on projects that are small and low impact.

The results depicted in Figure Table 7 suggest that availability of internal human resources supported organizational discussion and brainstorming. The grid again informed that such an activity was typically carried out for large and high profile projects. According to Table 8, those large/high-profile projects also led to IT Managers conducting market research. IT Managers further considered this approach to be solution and quality focused and enabled them to choose from different alternatives.

There is an even distribution of ratings in Table 9 in regard to the use of cost benefit analysis. This suggests that cost benefit analysis might be one of the more commonly employed, standard business practice activities to move projects forward from the initiation stage. However, it seems that this activity was carried out mainly for

Table 5. Analysis of ‘How-Why’ results; ‘Requirement Gathering’ element

| Requirement Gatherings | | | | |
|---|---|---|---|---|
| | 1s/2s | 3s | 5s/4s | |
| Faster development/implementation process | 1 | 0 | 6 | Quality focus |
| Clear responsibilities for all parties | 5 | 2 | 1 | Easy to use |
| Uncertainty of the solution/method | 2 | 1 | 0 | Availability of an instant solution |
| Time and/or budget constraints | 2 | 0 | 1 | Specified in process |
| Problem focused | 2 | 1 | 5 | Solution focused |
| Minimal risks | 0 | 3 | 1 | Governance procedures |
| Useful for communication | 2 | 2 | 3 | Useful for implementation |
| Small/low-impact project | 0 | 0 | 5 | Large/ high profile project |
| Easy to apply | 0 | 1 | 3 | Can be verified |
| Company's standard approach | 2 | 3 | 2 | Personal or team preference |
| The most common/acceptable practice in IT industry | 2 | 1 | 2 | A tailor-made process based on the project |
| Human resources constraints | 1 | 1 | 3 | Availability of internal staff/external consultant |
| Regulatory compliance | 0 | 1 | 4 | Freedom to choose |

As can be seen in Table 5, the requirement gathering technique is perceived as a quality and solution focused approach and the IT Managers that used it regarded it highly in terms of determining clear responsibilities for all parties. Interestingly, this approach was only seen as useful if applied in large/high profile projects. The reason may be large/high profile projects and perhaps to adhere to governance procedures.

As the results in Table 10 indicate, large and/or high profile projects again encouraged IT Managers to utilize a project management methodology. Despite the fact that IT

Table 6. Analysis of 'How-Why' results; 'No PM Method' element

| No PM Method | | | | |
|---|---|---|---|---|
| | 1s/2s | 3s | 5s/4s | |
| Faster development/implementation process | 1 | 0 | 0 | Quality focus |
| Clear responsibilities for all parties | 1 | 0 | 0 | Easy to use |
| Uncertainty of the solution/method | 2 | 0 | 0 | Availability of an instant solution |
| Time and/or budget constraints | 1 | 0 | 0 | Specified in process |
| Problem focused | 1 | 0 | 0 | Solution focused |
| Minimal risks | 0 | 1 | 0 | Governance procedures |
| Useful for communication | 0 | 1 | 0 | Useful for implementation |
| Small/low-impact project | 2 | 0 | 0 | Large/ high profile project |
| Easy to apply | 1 | 0 | 0 | Can be verified |
| Company's standard approach | 0 | 0 | 1 | Personal or team preference |
| The most common/acceptable practice in IT industry | 0 | 0 | 1 | A tailor-made process based on the project |
| Human resources constraints | 1 | 0 | 0 | Availability of internal staff/external consultant |
| Regulatory compliance | 0 | 0 | 1 | Freedom to choose |

Table 7. Analysis of 'How-Why' results; 'Internal Organization Discussion' element

| Internal Organisation Discussion | | | | |
|---|---|---|---|---|
| | 1s/2s | 3s | 5s/4s | |
| Faster development/implementation process | 1 | 2 | 2 | Quality focus |
| Clear responsibilities for all parties | 3 | 2 | 0 | Easy to use |
| Uncertainty of the solution/method | 3 | 3 | 0 | Availability of an instant solution |
| Time and/or budget constraints | 2 | 0 | 0 | Specified in process |
| Problem focused | 0 | 2 | 2 | Solution focused |
| Minimal risks | 0 | 3 | 0 | Governance procedures |
| Useful for communication | 3 | 2 | 1 | Useful for implementation |
| Small/low-impact project | 0 | 1 | 3 | Large/ high profile project |
| Easy to apply | 2 | 1 | 1 | Can be verified |
| Company's standard approach | 1 | 2 | 1 | Personal or team preference |
| The most common/acceptable practice in IT industry | 1 | 1 | 1 | A tailor-made process based on the project |
| Human resources constraints | 0 | 1 | 4 | Availability of internal staff/external consultant |
| Regulatory compliance | 0 | 0 | 3 | Freedom to choose |

Table 8. Analysis of 'How-Why' results; 'Market Research' element

| Market Research | | | | |
|---|---|---|---|---|
| | 1s/2s | 3s | 5s/4s | |
| Faster development/implementation process | 0 | 1 | 3 | Quality focus |
| Clear responsibilities for all parties | 1 | 1 | 1 | Easy to use |
| Uncertainty of the solution/method | 1 | 2 | 2 | Availability of an instant solution |
| Time and/or budget constraints | 3 | 0 | 0 | Specified in process |
| Problem focused | 0 | 0 | 5 | Solution focused |
| Minimal risks | 1 | 1 | 1 | Governance procedures |
| Useful for communication | 1 | 2 | 1 | Useful for implementation |
| Small/low-impact project | 0 | 0 | 6 | Large/ high profile project |
| Easy to apply | 2 | 2 | 1 | Can be verified |
| Company's standard approach | 1 | 2 | 1 | Personal or team preference |
| The most common/acceptable practice in IT industry | 2 | 2 | 0 | A tailor-made process based on the project |
| Human resources constraints | 0 | 3 | 0 | Availability of internal staff/external consultant |
| Regulatory compliance | 0 | 0 | 4 | Freedom to choose |

Table 9. Analysis of 'How-Why' results; 'Cost Benefit Analysis' element

| Cost Benefit Analysis | | | | |
|---|---|---|---|---|
| | 1s/2s | 3s | 5s/4s | |
| Faster development/implementation process | 0 | 1 | 2 | Quality focus |
| Clear responsibilities for all parties | 2 | 1 | 0 | Easy to use |
| Uncertainty of the solution/method | 1 | 1 | 0 | Availability of an instant solution |
| Time and/or budget constraints | 3 | 1 | 2 | Specified in process |
| Problem focused | 0 | 0 | 3 | Solution focused |
| Minimal risks | 1 | 1 | 3 | Governance procedures |
| Useful for communication | 1 | 2 | 0 | Useful for implementation |
| Small/low-impact project | 0 | 1 | 3 | Large/ high profile project |
| Easy to apply | 0 | 1 | 3 | Can be verified |
| Company's standard approach | 2 | 1 | 2 | Personal or team preference |
| The most common/acceptable practice in IT industry | 3 | 2 | 0 | A tailor-made process based on the project |
| Human resources constraints | 0 | 2 | 2 | Availability of internal staff/external consultant |
| Regulatory compliance | 1 | 0 | 3 | Freedom to choose |

Table 10. Analysis of 'How-Why' results; 'Project Management Methodology' element

| Project Management Methodology | | | | |
|---|---|---|---|---|
| | 1s/2s | 3s | 5s/4s | |
| Faster development/implementation process | 0 | 3 | 3 | Quality focus |
| Clear responsibilities for all parties | 4 | 2 | 0 | Easy to use |
| Uncertainty of the solution/method | 1 | 3 | 0 | Availability of an instant solution |
| Time and/or budget constraints | 4 | 1 | 0 | Specified in process |
| Problem focused | 0 | 2 | 4 | Solution focused |
| Minimal risks | 0 | 4 | 1 | Governance procedures |
| Useful for communication | 0 | 4 | 2 | Useful for implementation |
| Small/low-impact project | 0 | 0 | 5 | Large/ high profile project |
| Easy to apply | 1 | 1 | 4 | Can be verified |
| Company's standard approach | 1 | 3 | 3 | Personal or team preference |
| The most common/acceptable practice in IT industry | 4 | 1 | 0 | A tailor-made process based on the project |
| Human resources constraints | 0 | 1 | 3 | Availability of internal staff/external consultant |
| Regulatory compliance | 0 | 0 | 4 | Freedom to choose |

Table 11. Analysis of 'How-Why' results; 'Request for Proposal (RFP) Process' element

| Request for Proposal (RFP) Process | | | | |
|---|---|---|---|---|
| | 1s/2s | 3s | 5s/4s | |
| Faster development/implementation process | 0 | 1 | 3 | Quality focus |
| Clear responsibilities for all parties | 4 | 2 | 0 | Easy to use |
| Uncertainty of the solution/method | 2 | 2 | 0 | Availability of an instant solution |
| Time and/or budget constraints | 3 | 0 | 1 | Specified in process |
| Problem focused | 2 | 0 | 3 | Solution focused |
| Minimal risks | 0 | 3 | 2 | Governance procedures |
| Useful for communication | 1 | 4 | 0 | Useful for implementation |
| Small/low-impact project | 0 | 0 | 5 | Large/ high profile project |
| Easy to apply | 0 | 2 | 3 | Can be verified |
| Company's standard approach | 2 | 3 | 1 | Personal or team preference |
| The most common/acceptable practice in IT industry | 2 | 2 | 0 | A tailor-made process based on the project |
| Human resources constraints | 1 | 2 | 0 | Availability of internal staff/external consultant |
| Regulatory compliance | 1 | 1 | 3 | Freedom to choose |

Table 12. Analysis of 'How-Why' results; 'Company's Predefined Process' element

| Company's Predefined Process | | | | |
|---|---|---|---|---|
| | 1s/2s | 3s | 5s/4s | |
| Faster development/implementation process | 1 | 0 | 1 | Quality focus |
| Clear responsibilities for all parties | 1 | 1 | 0 | Easy to use |
| Uncertainty of the solution/method | 1 | 1 | 0 | Availability of an instant solution |
| Time and/or budget constraints | 2 | 0 | 0 | Specified in process |
| Problem focused | 1 | 0 | 1 | Solution focused |
| Minimal risks | 1 | 1 | 0 | Governance procedures |
| Useful for communication | 1 | 1 | 0 | Useful for implementation |
| Small/low-impact project | 1 | 0 | 1 | Large/ high profile project |
| Easy to apply | 1 | 1 | 0 | Can be verified |
| Company's standard approach | 2 | 0 | 1 | Personal or team preference |
| The most common/acceptable practice in IT industry | 0 | 1 | 1 | A tailor-made process based on the project |
| Human resources constraints | 1 | 1 | 0 | Availability of internal staff/external consultant |
| Regulatory compliance | 0 | 0 | 2 | Freedom to choose |

Table 13. Analysis of 'How-Why' results; 'Prototype' element

| Prototype | | | | |
|---|---|---|---|---|
| | 1s/2s | 3s | 5s/4s | |
| Faster development/implementation process | 2 | 2 | 1 | Quality focus |
| Clear responsibilities for all parties | 1 | 0 | 3 | Easy to use |
| Uncertainty of the solution/method | 3 | 1 | 1 | Availability of an instant solution |
| Time and/or budget constraints | 1 | 0 | 3 | Specified in process |
| Problem focused | 1 | 1 | 3 | Solution focused |
| Minimal risks | 4 | 2 | 0 | Governance procedures |
| Useful for communication | 1 | 3 | 1 | Useful for implementation |
| Small/low-impact project | 0 | 2 | 2 | Large/ high profile project |
| Easy to apply | 2 | 0 | 3 | Can be verified |
| Company's standard approach | 0 | 1 | 3 | Personal or team preference |
| The most common/acceptable practice in IT industry | 1 | 2 | 2 | A tailor-made process based on the project |
| Human resources constraints | 0 | 1 | 2 | Availability of internal staff/external consultant |
| Regulatory compliance | 0 | 0 | 3 | Freedom to choose |

Table 14. Analysis of ‘How-Why’ results; ‘Site Visits’ element

| Site Visits | | | | |
|---|---|---|---|---|
| | 1s/2s | 3s | 5s/4s | |
| Faster development/implementation process | 1 | 1 | 1 | Quality focus |
| Clear responsibilities for all parties | 0 | 0 | 1 | Easy to use |
| Uncertainty of the solution/method | 1 | 1 | 0 | Availability of an instant solution |
| Time and/or budget constraints | 1 | 0 | 0 | Specified in process |
| Problem focused | 0 | 0 | 2 | Solution focused |
| Minimal risks | 0 | 1 | 0 | Governance procedures |
| Useful for communication | 0 | 1 | 0 | Useful for implementation |
| Small/low-impact project | 0 | 0 | 2 | Large/ high profile project |
| Easy to apply | 0 | 1 | 1 | Can be verified |
| Company's standard approach | 0 | 1 | 0 | Personal or team preference |
| The most common/acceptable practice in IT industry | 1 | 1 | 0 | A tailor-made process based on the project |
| Human resources constraints | 0 | 1 | 0 | Availability of internal staff/external consultant |
| Regulatory compliance | 0 | 0 | 1 | Freedom to choose |

Table 15. Analysis of ‘How-Why’ results; ‘Evaluative Framework’ element

| Evaluative Framework | | | | |
|---|---|---|---|---|
| | 1s/2s | 3s | 5s/4s | |
| Faster development/implementation process | 0 | 0 | 1 | Quality focus |
| Clear responsibilities for all parties | 1 | 0 | 0 | Easy to use |
| Uncertainty of the solution/method | 0 | 0 | 0 | Availability of an instant solution |
| Time and/or budget constraints | 0 | 0 | 0 | Specified in process |
| Problem focused | 0 | 2 | 0 | Solution focused |
| Minimal risks | 1 | 1 | 0 | Governance procedures |
| Useful for communication | 2 | 1 | 0 | Useful for implementation |
| Small/low-impact project | 0 | 0 | 1 | Large/ high profile project |
| Easy to apply | 1 | 0 | 1 | Can be verified |
| Company's standard approach | 0 | 1 | 1 | Personal or team preference |
| The most common/acceptable practice in IT industry | 1 | 0 | 0 | A tailor-made process based on the project |
| Human resources constraints | 0 | 0 | 0 | Availability of internal staff/external consultant |
| Regulatory compliance | 0 | 0 | 1 | Freedom to choose |

Table 16. Analysis of ‘How-Why’ results; ‘Narrative Specs’ element

| Narrative Specs | | | | |
|---|---|---|---|---|
| | 1s/2s | 3s | 5s/4s | |
| Faster development/implementation process | 0 | 1 | 1 | Quality focus |
| Clear responsibilities for all parties | 0 | 1 | 0 | Easy to use |
| Uncertainty of the solution/method | 0 | 2 | 0 | Availability of an instant solution |
| Time and/or budget constraints | 1 | 0 | 0 | Specified in process |
| Problem focused | 0 | 1 | 1 | Solution focused |
| Minimal risks | 0 | 2 | 0 | Governance procedures |
| Useful for communication | 0 | 2 | 0 | Useful for implementation |
| Small/low-impact project | 0 | 0 | 2 | Large/ high profile project |
| Easy to apply | 0 | 0 | 2 | Can be verified |
| Company's standard approach | 0 | 2 | 0 | Personal or team preference |
| The most common/acceptable practice in IT industry | 1 | 1 | 0 | A tailor-made process based on the project |
| Human resources constraints | 0 | 1 | 0 | Availability of internal staff/external consultant |
| Regulatory compliance | 0 | 0 | 1 | Freedom to choose |

Table 17. Analysis of ‘How-Why’ results; ‘Models’ element

| Models | | | | |
|---|---|---|---|---|
| | 1s/2s | 3s | 5s/4s | |
| Faster development/implementation process | 0 | 1 | 2 | Quality focus |
| Clear responsibilities for all parties | 0 | 1 | 1 | Easy to use |
| Uncertainty of the solution/method | 1 | 2 | 0 | Availability of an instant solution |
| Time and/or budget constraints | 1 | 0 | 1 | Specified in process |
| Problem focused | 0 | 0 | 2 | Solution focused |
| Minimal risks | 0 | 1 | 1 | Governance procedures |
| Useful for communication | 1 | 3 | 0 | Useful for implementation |
| Small/low-impact project | 0 | 0 | 3 | Large/ high profile project |
| Easy to apply | 0 | 1 | 2 | Can be verified |
| Company's standard approach | 0 | 2 | 1 | Personal or team preference |
| The most common/acceptable practice in IT industry | 2 | 1 | 0 | A tailor-made process based on the project |
| Human resources constraints | 0 | 1 | 1 | Availability of internal staff/external consultant |
| Regulatory compliance | 1 | 0 | 1 | Freedom to choose |

Managers did not consider the application of a project management methodology as an easy process, they all regarded the process as useful, accountable, responsible, verifiable and of value.

Similar to previous grids, IT Managers carried out a request for proposal (RFP) process when projects were large and/or high profile. RFPs were seen as useful for both communication and for implementation, and helped in the delineation of responsibilities (as shown in Table 11).

It appears (Table 12) that there were not many projects that required IT Managers to follow the organization's pre-defined process (or might indicate that such pre-defined processes simply did not exist).

The results shown in Table 13 suggest that there are different perceptions among IT Managers regarding the reasons for using prototypes. However, one aspect that did generate agreement is the role of prototypes in minimizing risks.

Based on the collected data presented in Table 14, in spite of some support the site visit was not a commonly used method to advance a project from the initiation stage.

As can be seen in Tables 15 to 18, the construct ratings for the evaluative framework and narrative specs approaches are widely spread and there are no common reasons for performing these activities, on the rare occasions that they were used. Likewise, Models and an RFI process were used for a range of reasons. However, one main theme in applying these approaches was, again, their particular utility in large and/or high profile projects.

***Discussion***

In summarizing the above findings, it appears that IT Managers tended to use the more commonly recommended approaches (i.e., those recommended by the practitioner and research communities) principally in large and/or high profile projects. For example, requirements gathering/analysis, a project management methodology, vendor's demos, market research and RFP processes received general support. The findings further suggest that IT Managers are generally aware of the requirements of such processes/activities and the benefits of adopting these approaches and methods. However, the results suggest that smaller projects are treated with less formal processes. On the other hand, some IT Managers/organizations do not apply these common approaches at all, in any of their projects.

## 7. CONCLUSIONS

There are a number of key findings emanating from this research. With regard to the 'Why' question – Why Are ICT Projects Undertaken? – inefficient systems or processes was reported as the most influential factor encouraging IT Managers to initiate a project. This was followed by the drive for cost savings, the second most highly-rated rationale that influenced projects being proposed. In many respects these two reasons represent established bases for innovation – the desire to become more efficient or less expensive in operation have long motivated organizational change. In relation to the 'How' question – How Did You Move To A Solution? – cost-benefit analysis was the most favored approach used. The sample of IT Managers interviewed here rated requirements gathering and analysis as the second-top approach. Project management methodologies and internal (within-organization) discussions were also employed relatively often. The IT Managers appeared to utilize specific approaches for large and/or high profile projects.

The relatively poor record of IT project success over a long period of time suggests that IT Managers need to be highly cautious in making project decisions. As new technologies continue to emerge so new project opportunities will arise. Therefore, close attention is required to evaluate the motivating factors underpinning project initiation decisions. Such decisions, given their potential consequences, need to be justifiable, transparent and auditable. While it may be true that technology/systems development is in many respects an intangible/unquantifiable intellectual property of an organization, it is equally true that a successful project can deliver tremendous organizational benefits and/or cost savings, at a potentially substantial cost. Therefore, this intangibility should not be used to excuse an organization from systematically and comprehensively addressing the rationale for their IT projects. In particular, those projects that have the potential to deliver the most value are also often those that are the most challenging, and therefore the most costly. The rationale for project initiation decisions not only deserve a thorough evaluation at that time, but should also be revisited as part of the monitoring process at the post-project stage.

Table 18. Analysis of 'How-Why' results; 'RFI Process' element

| Request for Information (RFI) Process | | | | |
|---|---|---|---|---|
| | 1s/2s | 3s | 5s/4s | |
| Faster development/implementation process | 0 | 0 | 2 | Quality focus |
| Clear responsibilities for all parties | 1 | 0 | 0 | Easy to use |
| Uncertainty of the solution/method | 1 | 1 | 0 | Availability of an instant solution |
| Time and/or budget constraints | 0 | 0 | 1 | Specified in process |
| Problem focused | 0 | 0 | 1 | Solution focused |
| Minimal risks | 0 | 1 | 1 | Governance procedures |
| Useful for communication | 1 | 1 | 0 | Useful for implementation |
| Small/low-impact project | 0 | 0 | 2 | Large/ high profile project |
| Easy to apply | 0 | 1 | 1 | Can be verified |
| Company's standard approach | 1 | 1 | 0 | Personal or team preference |
| The most common/acceptable practice in IT industry | 1 | 1 | 0 | A tailor-made process based on the project |
| Human resources constraints | 0 | 1 | 0 | Availability of internal staff/external consultant |
| Regulatory compliance | 0 | 1 | 0 | Freedom to choose |

The potential for further research as a consequence of this study could follow several directions. In-depth case study analyses of individual projects in multiple organizations are likely to contribute greater understanding of project-specific motivating factors and decision patterns. The underlying reasons and decision patterns across specific project categories (such as system implementation or software development projects, infrastructure-related projects, system/process improvement projects, integration projects and Business Intelligence projects) may also be another research direction. With regard to the employment of the customized research methodology in this research, a study on the effectiveness and advantages/disadvantages of a tailor-made RepGrid method in other studies could be conducted.

## 8. ACKNOWLEDGMENTS

We acknowledge with gratitude the vital contributions made by the study participants.

## REFERENCES

Alexander, P., Loggerenberg, J. V., Lotriet, H., & Phahlamohlaka, J. (2010). The use of the Repertory Grid for collaboration and reflection in a research context. Group Decision and Negotiation, 19(5), 479-504.

Aurum, A., & Wohlin, C. (2005). Engineering and Managing Software Requirements. New York: Springer.

Cadle, J., & Yeates, D. (2008). Project Management for Information Systems (Fifth ed.). London: Pearson Education Limited.

Edwards, H. M., McDonald, S., & Young, S. M. (2009). The repertory grid technique: Its place in empirical software engineering research. Information and Software Technology, 51(4), 785- 798.

Guah, M. W. (2008). IT project escalation: A case analysis within the UK NHS. International Journal of Information Management, 28(6), 536-540.

Hansen, S., Berente, N., & Lyytinen, K. (2009). Requirements in the 21st Century: Current practice and emerging trends, in Design Requirements Engineering: A Ten-Year Perspective. (K. Lyytinen, J. Loucopoulos, J. Mylopoulos, & B. Robinson, Eds.) LNBIP 14, Springer, pp. 44-87.

Hillson, D. (2013). Resolving Cobb's paradox, Risk Management Today, 30, 181-182.

Hunter, M. G., & Beck, J. E. (2000). Using repertory grids to conduct cross-cultural information systems research. Information Systems Research, 11(1), 93-101.

Jani, A. (2008). An experimental investigation of factors influencing perceived control over a failing IT project. International Journal of Project Management, 26, 726-732.

Keil, M., Wallace, L., Turk, D., Dixon-Randall, G., & Nulden, U. (2000). An investigation of risk perception and risk propensity on the decision to continue a software development project. The Journal of Systems and Software, 53, 145-157.

Kelly, G. A. (1955). A theory of personality: The psychology of personal constructs. New York: W.W. Norton.

Napier, N. P., Keil, M., & Tan, F. B. (2009). IT project mangers' construction of successful project management practice: a repertory grid investigation. Information Systems Journal, 19(3), 255-282.

Rognerud, H. J., & Hannay, J. E. (2009). Challenges in enterprise software integration: An industrial study using Repertory Grids. Third Symposium on Empirical Software Engineering and Management, IEEE CS Press, 11-22.

Sauer, C., Gemino, A., & Reich, B. H. (2007). The impact of size and volatility on IT project performance. Communications of the ACM, 50(11), 79-84.

Seiler, S., Lent, B., Pinkowska, M., & Pinazza, M. (2012). An integrated model of factors influencing project managers' motivation - Findings from a Swiss survey. International Journal of Project Management, 30, 60-72.

Shim, S., Chae, M., & Lee, B. (2009). Empirical analysis of risk-taking behavior in IT platform migration decisions. Computers in Human Behaviour, 25, 1290-1305.

Siau, K., Tan, X., & Sheng, H. (2010). Important characteristics of software development team members: An empirical investigation using Repertory Grid. Information Systems Journal, 20(6), 563-580.

Sommerville, I., Cliff, D., Calinescu, R., Keen, J., Kelly, T., Kwiatkowska, M., McDermid, J., & Paige, R. (2012). Large-scale complex IT systems, Communications of the ACM, 55(7), 71-77.

Tan, F. B., & Hunter, M. G. (2002). The Repertory Grid technique: A method for the study of cognition in information systems. MIS Quarterly, 26(1), 39-57.